      [1999/12/01 v1.4c Il Nuovo Cimento]
\documentclass[preprint]{cimento}
\usepackage{overpic}
\usepackage{booktabs,tabulary}
\usepackage{cite}
\usepackage[utf8]{inputenc}
\usepackage{cancel}
\usepackage{amsmath}
\usepackage{multirow}
\usepackage{graphicx}
\newcommand{\eq}[1]{Eq.~(\ref{#1})}
\newcommand{\ps}{p\hspace{-0.44em}/\hspace{0.06em}}
\newcommand{\Msusy}{M_{\textrm{SUSY}}}

\newcommand{\gev}{\, \mathrm{GeV}}
\newcommand{\tev}{\, \mathrm{TeV}}

\newcommand{\Lagr}{\mathcal{L}}
\newcommand{\beq}{\begin{equation}}
\newcommand{\eeq}{\end{equation}}
\renewcommand{\Re}{\text{Re}\,}
\renewcommand{\Im}{\text{Im}\,}

\allowdisplaybreaks[1]

\title{The Anomalous Magnetic Moment of the Muon:\\ Beyond the Standard Model via Chiral Enhancement}
\shorttitle{The AMM of the muon:  Beyond the Standard Model via Chiral Enhancement}
\author{Andreas Crivellin\from{ins:x}\from{ins:y} and Martin Hoferichter\from{ins:z}}
\instlist{\inst{ins:x} Paul Scherrer Institut, 5232 Villigen PSI, Switzerland
\inst{ins:y}Physik-Institut, Universit\"at Z\"urich, Winterthurerstrasse 190, 8057 Z\"urich, Switzerland
\inst{ins:z} Albert Einstein Center for Fundamental Physics, Institute for Theoretical Physics, University of Bern, Sidlerstrasse 5, 3012 Bern, Switzerland}
\begin{document}
\maketitle
\begin{abstract}
The anomalous magnetic moment of the muon displays a $4.2\sigma$ tension with the Standard-Model prediction, if $e^+e^-\to \text{hadrons}$ data are used for hadronic vacuum polarization. In these proceedings we review possible explanations of this anomaly in terms of heavy new particles. As the necessary effect is of the order of the electroweak Standard-Model contribution, viable explanations with TeV-scale physics must involve an enhancement factor; in particular, one can obtain the chirality flip of the dipole operator via a sizable coupling to the Higgs doublet instead of the small muon Yukawa coupling. Such Standard-Model extensions then also predict effects in Higgs and $Z$-boson decays to muons, with details depending on the $SU(2)_L\times U(1)_Y$ representations of the new particles. We first review the general case of chirally enhanced new physics, before discussing in more detail the concrete example of leptoquark models. 
\end{abstract}

\section{Introduction}

The Run~1 result of the Fermilab Muon $g-2$ experiment~\cite{Abi:2021gix} confirmed the previous measurement at Brookhaven National Laboratory~\cite{Bennett:2006fi}, leading to a combined world average of 
\begin{equation}
a_\mu^\text{exp}=116\,592\,061(41)\times 10^{-11},
\end{equation}
which differs from the Standard-Model (SM) theory prediction~\cite{Aoyama:2020ynm} (mainly based on Refs.~\cite{Aoyama:2012wk,Aoyama:2019ryr,Czarnecki:2002nt,Gnendiger:2013pva,Davier:2017zfy,Keshavarzi:2018mgv,Colangelo:2018mtw,Hoferichter:2019mqg,Davier:2019can,Keshavarzi:2019abf,Kurz:2014wya,Melnikov:2003xd,Masjuan:2017tvw,Colangelo:2017fiz,Hoferichter:2018kwz,Gerardin:2019vio,Bijnens:2019ghy,Colangelo:2019uex,Blum:2019ugy,Colangelo:2014qya})
\begin{equation}
\label{amuSM}
a_\mu^\text{SM}=116\,591\,810(43)\times 10^{-11},
\end{equation}
by $4.2\sigma$.\footnote{The lattice-QCD results for hadronic vacuum polarization~\cite{Borsanyi:2020mff,Ce:2022kxy,Alexandrou:2022amy} indicate a smaller tension than the SM prediction in Ref.~\cite{Aoyama:2020ynm}, but display a tension with $e^+e^-$ data that for the intermediate region in center-of-mass energy reaches $4\sigma$~\cite{Borsanyi:2020mff,Ce:2022kxy,Alexandrou:2022amy,Colangelo:2022vok}, see also Refs.~\cite{Crivellin:2020zul,Keshavarzi:2020bfy,Malaescu:2020zuc,Colangelo:2020lcg} for the consequences of this emerging tension. Note, however, that the lattice-QCD calculation of Ref.~\cite{Blum:2018mom} agrees with $e^+e^-$ data for this intermediate ``window'' quantity, and further calculations are required to draw firm conclusions.} If this tension is indeed a signal of physics beyond the SM, the most pressing challenge is unraveling its nature. 

Given that $\Delta a_\mu=a_\mu^\text{exp}-a_\mu^\text{SM}=251(59)\times 10^{-11}$ is larger than the electroweak (EW) contribution of the SM, $a_\mu^\text{EW}=153.6(1.0)\times 10^{-11}$~\cite{Czarnecki:2002nt,Gnendiger:2013pva}, any BSM explanation must be able to provide such a sizable effect. Apart from light new physics, which we will not consider here (see Ref.~\cite{Athron:2021iuf} for an extensive review of new physics in $(g-2)_\mu$), this can be achieved by a chirality flip originating from a large coupling to the SM Higgs instead of the small muon Yukawa coupling in the SM. This chiral enhancement allows for viable solutions for particle masses up to tens of TeV~\cite{Czarnecki:2001pv,Stockinger:1900zz,Giudice:2012ms,Altmannshofer:2016oaq,Kowalska:2017iqv,Crivellin:2018qmi,Capdevilla:2021rwo}. However,
the same mechanism also produces sizable effects in other processes, in particular $h\to\mu^+\mu^-$ and $Z\to\mu^+\mu^-$, albeit in general not probing the exact same combination of couplings. These correlations are therefore not model independent, but require at least the specification of the $SU(2)_L\times U(1)_Y$ representations of the new particles. Here, we first review the general argument, before presenting concrete examples in leptoquark (LQ) models.

\section{Generic New Physics Effects}

The effective Hamiltonian governing dipole transitions of charged leptons is given by
\beq
\label{HeffLFV}
{\cal{H}}_\text{eff}= c^{\ell_{f}\ell_{i}}_{R} \, \bar{\ell}_{f}
\sigma_{\mu \nu} P_{R} \ell_{i} F^{\mu \nu}+\text{h.c.},
\eeq
with
\begin{align}
\begin{aligned}
\label{Brmuegamma}
a_{\ell_i}&=\frac{(g-2)_{\ell_i}}{2}= -\frac{2m_{\ell_{i}}}{e}\, \big(c^{\ell_{i}\ell_{i}}_{R}+c^{\ell_{i}\ell_{i}*}_{R}\big)=-\frac{4m_{\ell_{i}}}{e}\, \Re c^{\ell_{i}\ell_{i}}_{R},\\
d_{\ell_i} &= i\big(c^{\ell_{i}\ell_{i}}_{R}-c^{\ell_{i}\ell_{i}*}_{R}\big)=-2\,\Im c^{\ell_{i}\ell_{i}}_{R},
\end{aligned}
\end{align}
where $\ell_i,\ell_f\in\{e,\mu,\tau\}$. This emphasizes that the Wilson coefficients~\eqref{HeffLFV} are not necessarily subject to minimal flavor violation, i.e., $a_\ell$ is linear (rather than quadratic) in $m_\ell$ and the phase of the Wilson coefficient for each flavor can be different.

In general, the dipole operator~\eqref{HeffLFV} does not imply immediate correlations with other processes, but such connections can still be established for a wide range of simplified models, i.e., we can parameterize the couplings of new scalars/vectors to SM leptons  $\ell_i$ contributing to the dipole operators as
\begin{align}
\begin{aligned}
\label{Lagr_general}
\Lagr_\Phi&= \bar \Psi \left( {\Gamma _{\Psi \Phi }^{iL}{P_L} + \Gamma _{\Psi \Phi }^{iR}{P_R}} \right){\ell _i}{\Phi ^*} +\text{h.c.},\\
\Lagr_{V}&= \bar \Psi \left( {\Gamma _{\Psi V }^{iL}\gamma^\mu{P_L} + \Gamma _{\Psi V }^{iR}\gamma^\mu{P_R}} \right){\ell _i}{V_\mu ^*} +\text{h.c.},
\end{aligned}
\end{align}
where a sum over all fermions $\Psi$ and scalars (vectors) $\Phi$ ($V^\mu$) is implicitly understood. With these conventions at hand the contribution to the Wilson coefficients (for heavy new physics) are~\cite{Crivellin:2018qmi} (see also Refs.~\cite{Freitas:2014pua,Stockinger:2006zn})
\begin{align}
c_{R\Phi}^{fi} &= \frac{e}{{16{\pi ^2}}}\Gamma _{\Psi \Phi }^{fL*}\Gamma _{\Psi \Phi }^{iR}{M_\Psi }\frac{f_\Phi\big(\frac{M_\Psi^2}{M_\Phi^2}\big)+Q g_\Phi\big(\frac{M_\Psi^2}{M_\Phi^2}\big)}{M_\Phi^2}\label{c_R_general_Phi}\\
&+ \frac{e}{{16{\pi ^2}}}\big(m_{\ell_i}\Gamma _{\Psi\Phi }^{fL*}\Gamma _{\Psi\Phi }^{iL}+m_{\ell_f}\Gamma _{\Psi\Phi }^{fR*}\Gamma _{\Psi\Phi }^{iR}\big)\times\frac{\tilde f_\Phi\big(\frac{M_\Psi^2}{M_\Phi^2}\big)+Q \tilde g_\Phi\big(\frac{M_\Psi^2}{M_\Phi^2}\big)}{M_\Phi^2} ,\notag\\
c_{R V}^{fi} &= \frac{e}{{16{\pi ^2}}}\Gamma _{\Psi V }^{fL*}\Gamma _{\Psi V }^{iR}M_\Psi \frac{{{f_V}\big(\frac{M_\Psi^2}{M_V^2}\big) + Q{g_V}\big(\frac{M_\Psi^2}{M_V^2}\big)}}{{M_V ^2}}\label{c_R_general_V}\\
&+\frac{e}{{16{\pi ^2}}}\big(m_{\ell_i}\Gamma _{\Psi V}^{fL*}\Gamma _{\Psi V}^{iL}+m_{\ell_f}\Gamma _{\Psi V}^{fR*}\Gamma _{\Psi V}^{iR}\big)\times\frac{{{\tilde f_V}\big(\frac{M_\Psi^2}{M_V^2}\big) + Q{\tilde g_V}\big(\frac{M_\Psi^2}{M_V^2}\big)}}{{M_V ^2}},\notag
\end{align}
with loop functions
\begin{align}
\begin{aligned}
f_\Phi(x)&=2\tilde g_\Phi(x)=\frac{x^2-1-2x\log x}{4(x-1)^3},\\
g_\Phi(x)&=\frac{x-1-\log x}{2(x-1)^2},\\
\tilde f_\Phi(x)&=\frac{2x^3+3x^2-6x+1-6x^2\log x}{24(x-1)^4},\\
f_V(x) &= \frac{x^3-12x^2 + 15x -4 + 6x^2\log x}{4(x - 1)^3},\\
g_V(x) &= \frac{x^2 - 5x +4 + 3x\log x}{2(x - 1)^2},\\
\tilde f_V(x) &= \frac{-4x^4+49x^3-78x^2 + 43x -10 - 18x^3\log x}{24(x - 1)^4},\\
\tilde g_V(x) &= \frac{-3(x^3-6x^2 + 7x -2 + 2x^2\log x)}{8(x - 1)^3},\label{loop_functions}
\end{aligned}
\end{align}
where $Q$ is the electric charge of the fermion. We calculated the contribution of the massive vector boson in unitary gauge, so that the effects of Goldstone bosons are automatically included, which is possible since the matching on dipole operators gives a finite result. The terms proportional to the heavy fermion mass are the ones that can be chirally enhanced. These contributions have an arbitrary phase also for $i = f$ while, due to Hermiticity of the Lagrangian, the terms that are not chirally enhanced, i.e., proportional to $\Gamma _{\Psi V,\Phi }^{fL*}\Gamma _{\Psi V,\Phi }^{iL}$ and $\Gamma _{\Psi V,\Phi }^{fR*}\Gamma _{\Psi V,\Phi }^{iR}$ (included here for completeness), are real for flavor-conserving dipole transitions. Importantly, this means that in case of a chirally enhanced new physics effect, the Wilson coefficients of the corresponding dipole operators are in general complex.  While the real part generates an effect in the anomalous magnetic moment, the imaginary part is related to the electric dipole moment of the muon, and absent further restrictions can thus be sizable~\cite{Crivellin:2018qmi,Crivellin:2019mvj}, potentially within reach of future experiments~\cite{Adelmann:2021udj,Aiba:2021bxe}.

\begin{table}[t!]
	\centering
	\begin{tabular}{cc|ccc|ccc}
		\toprule
		& $R$ & ${\Psi ,\Phi }$&${{\Phi _L},{\Psi _L}}$&${{\Phi _E},{\Psi _E}}$ & $\phi$ & $\ell$ & $e$\\\midrule
		\multirow{4}{*}{$SU(2)_L$} & $121$ & $1$ & $2$ & $1$ & \multirow{4}{*}{$2$} & \multirow{4}{*}{$2$} & \multirow{4}{*}{$1$}\\
		& $212$ & $2$ & $1$ & $2$ & & &\\
		& $323$ & $3$ & $2$ & $3$ & & &\\
		& $232$ & $2$ & $3$ & $2$ & & &\\
		$Y$ & & $X$ & $-\frac{1}{2}-X$ & $-1-X$ & $\frac{1}{2}$ & $-\frac{1}{2}$ &$-1$\\
		\bottomrule
	\end{tabular}
	\caption{Charge assignments and representations under $SU(2)_L\times U(1)_Y$ for the SM fields and the different new particles. Hypercharge is parameterized by a free parameter $X$.}
	\label{quantum_numbers}
\end{table}

\begin{figure}[t!]
	\centering
	\includegraphics[width=0.39\textwidth]{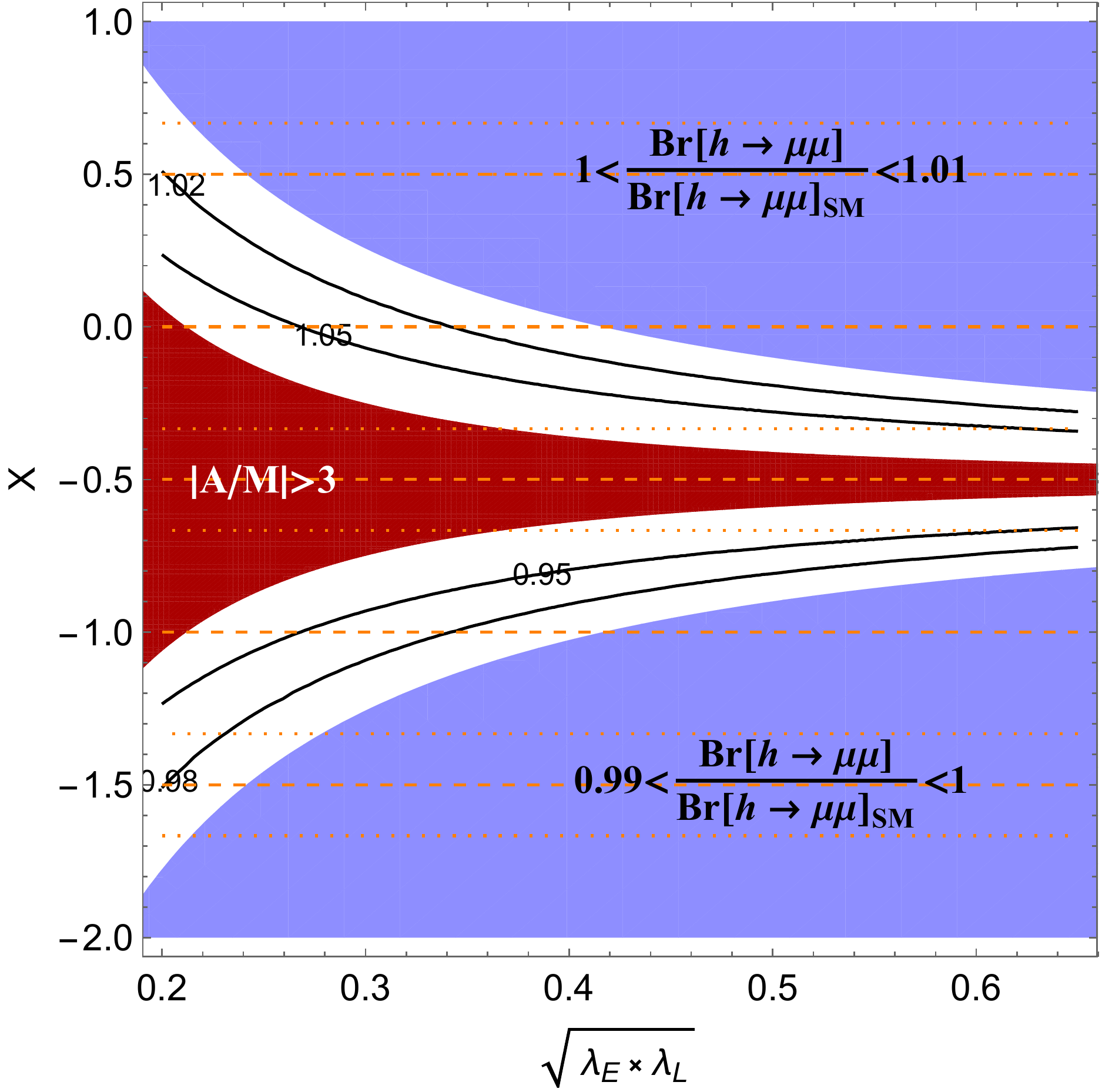}\qquad
	\includegraphics[width=0.39\textwidth]{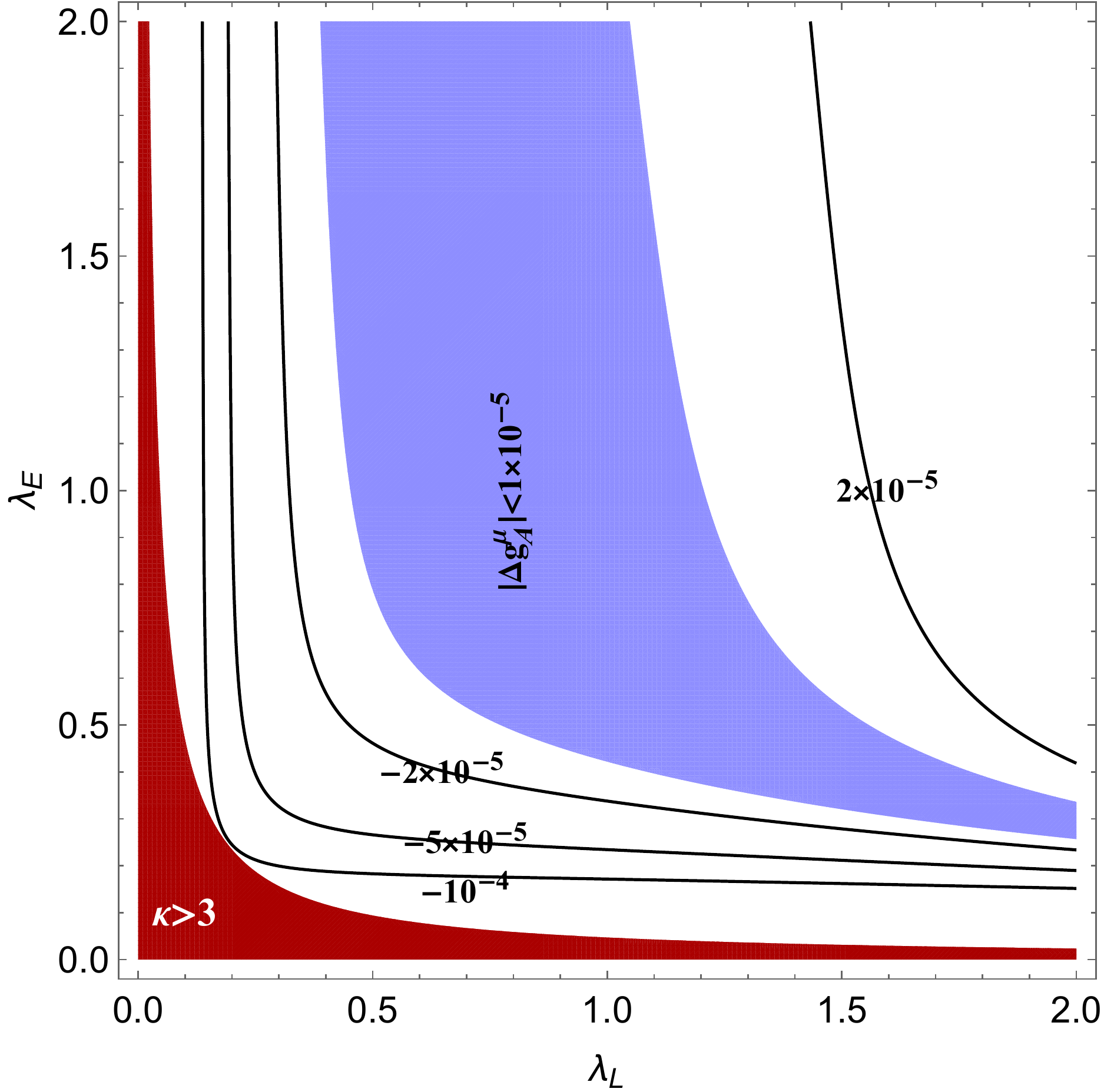}
	\caption{Effects in $h\to\mu^+\mu^-$ (left, case I with $R=121$) and $Z\to\mu^+\mu^-$ (right, case II with $R=232$), when imposing the constraint to reproduce $\Delta a_\mu$. In both cases, in the blue regions the effect is too small to be observable, in the red region the couplings become non-perturbative, but the parameter space in between can be probed at future colliders. Taken from Ref.~\cite{Crivellin:2021rbq}.}
	\label{fig:chiral}
\end{figure}

\section{$\boldsymbol{SU(2)_L}$ invariance}

The relations in Eqs.~\eqref{Lagr_general}--\eqref{c_R_general_V} are not manifestly $SU(2)_L$ invariant, but only invariant with respect to $U(1)_\text{EM}$. However, for new physics realized above the EW scale, $SU(2)_L$ gauge symmetry must be respected. Therefore, let us consider a class of models with new scalars and fermions that display the minimal features required to implement chiral enhancement, allowing for a wide range of $SU(2)_L$ representations and hypercharges. We can match these models onto the relevant set of dimension-$6$ effective operators in SMEFT~\cite{Buchmuller:1985jz,Grzadkowski:2010es}, based on which correlations have been pointed out in Refs.~\cite{Buttazzo:2020eyl,Yin:2020afe,Aebischer:2021uvt}. 

In order to achieve chiral enhancement in $a_\mu$ with new particles in the loop, at least three fields (two scalars and one fermion or two fermions and one scalar) are needed, some of which, as, e.g., in LQ models, can be taken from the SM. As a first step, 
we classify the possible representations of three new fields under $SU(2)_L\times U(1)_Y$ and perform the matching onto the relevant SMEFT operators. 

There are two classes of models that display chiral enhancement for $a_\mu$: (I)  two scalars $\Phi_{L,E}$ and one fermion $\Psi$ and  
(II) two fermions $\Psi_{L,E}$ and one scalar $\Phi$. We thus define the Lagrangians in these two cases as
\begin{align}
\begin{aligned}
{\Lagr_\text{I}} &= \lambda _L^{\text{I}}\,\bar \ell\Psi {\Phi _L} + \lambda _E^{\text{I}}\,\bar e\Psi {\Phi _E} + A\,\Phi _L^\dag \Phi _E^{}\phi,\\
{\Lagr_\text{II}} &= \lambda _L^\text{II}\,\bar \ell{\Psi _L}\Phi  + \lambda _E^\text{II}\,\bar e{\Psi _E}\Phi  + \kappa\, {{\bar \Psi }_L}{\Psi _E}\phi,
\end{aligned}
\end{align}
where $\ell$, $e$, and $\phi$ are the lepton doublet, singlet, and Higgs field of the SM (throughout, we follow the notation of Ref.~\cite{Grzadkowski:2010es}). The conventions for the SM particles and the $SU(2)_L$ quantum numbers and hypercharges $Y$ are given in Table~\ref{quantum_numbers}. We consider the four combinations of $SU(2)_L$ representations ($R$) up-to-and-including triplets (see also Ref.~\cite{Calibbi:2018rzv}), while the hypercharge assignment can be parameterized in terms of a general variable $X$. We further assume a $Z_2$ symmetry to avoid mixing with SM fields, which could generate tree-level effects in $h\to\mu^+\mu^-$ and $Z\to\mu^+\mu^-$ that are at least strongly disfavored~\cite{ALEPH:2005ab,Crivellin:2018qmi,Zyla:2020zbs,Crivellin:2020ebi,Aad:2020xfq,Sirunyan:2020two,Crivellin:2020tsz}. Moreover, $a_\mu$ is only generated at loop level, further motivating the study of loop effects in $h\to\mu^+ \mu^-$ and $Z\to\mu^+\mu^-$ as well. The full results for all cases discussed above can be found in Ref.~\cite{Crivellin:2021rbq}, in Fig.~\ref{fig:chiral} we show two examples for the numerical analysis. In particular, the general case also covers the LQ models $S_1$ and $S_2$ within case II with $R=121$ and $212$, respectively, and upon identifying $\Psi_L=t_L$, $\Psi_E=t_R$, and  $\kappa=Y_t$, which is the special case we discuss in more detail in the following section.

\begin{figure}[t]
	\centering
	\includegraphics[width=0.59\textwidth]{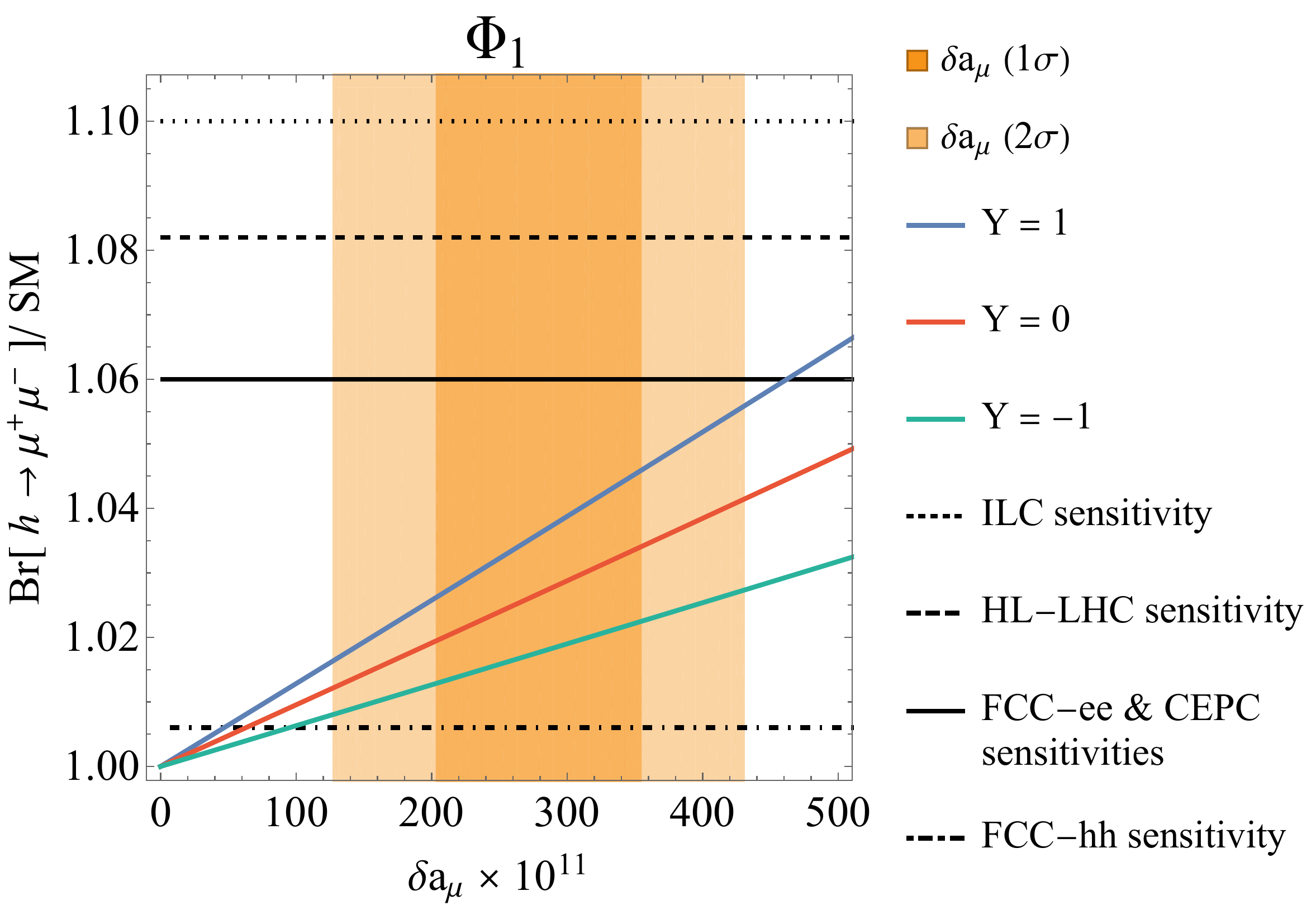}
	\includegraphics[width=0.395\textwidth]{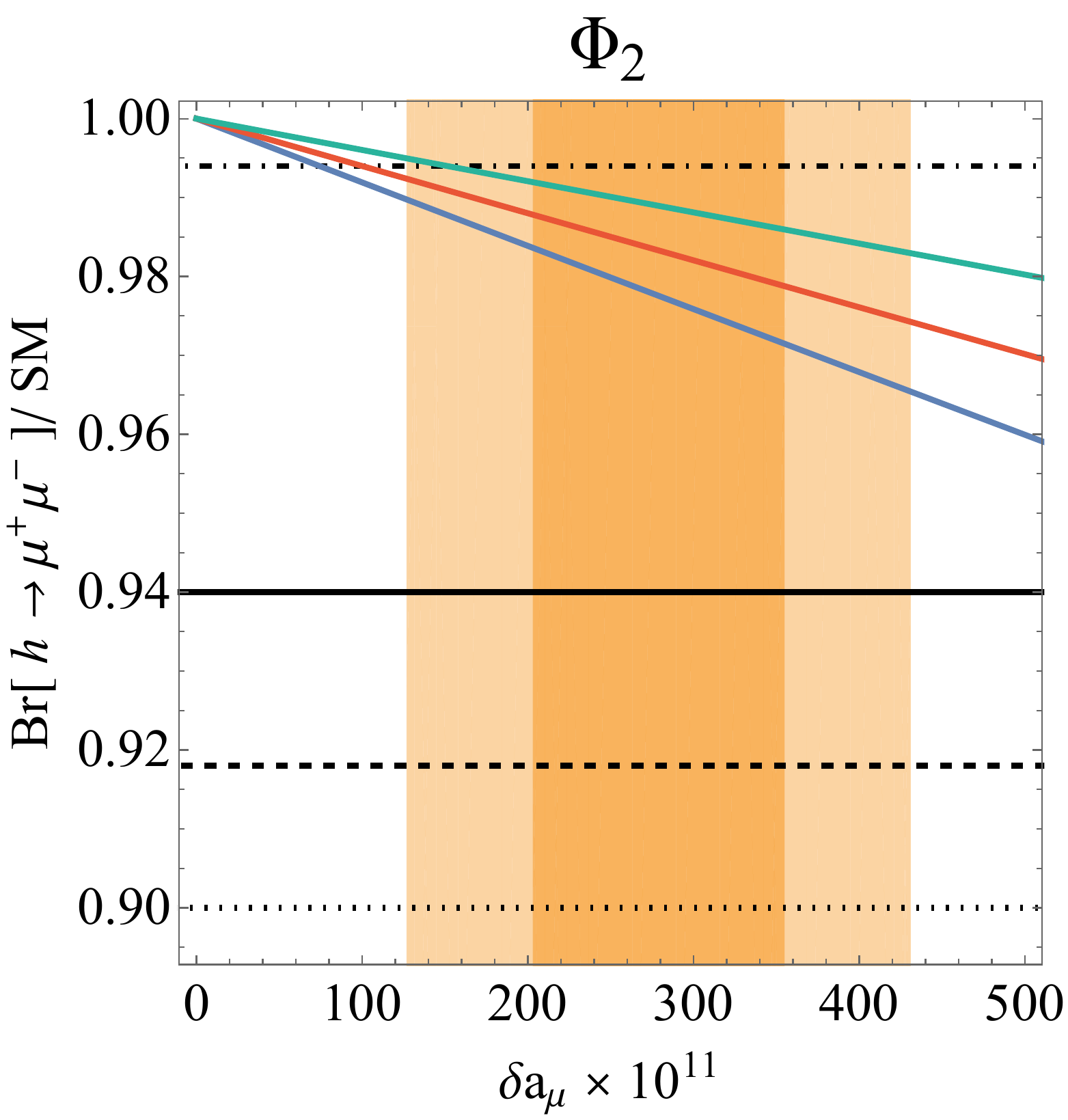}
	\caption{Correlations between ${\rm Br}[h\to\mu^+\mu^-]$, normalized to its SM value, and the new-physics contribution to $a_\mu$ ($\delta a_\mu$) for scenario $\Phi_1$ (left) and $\Phi_2$ (right) with \mbox{$m_{1,2}=1.5\,$TeV}. The predictions for different values of the LQ couplings to the Higgs are shown, where for $\Phi_1$ $Y=Y_1$ and for $\Phi_2$ $Y=Y_2+Y_{22}$. Even though the current ATLAS and CMS results are not yet constraining this model, sizable effects are predicted, which can be tested at future colliders. Furthermore, $\Phi_1$ yields a constructive effect in $h\to\mu^+\mu^-$, while the one of $\Phi_2$ is destructive, such that they can be clearly distinguished with increasing experimental precision. Taken from Ref.~\cite{Crivellin:2020tsz}.}
	\label{S1S2}
\end{figure}

\section{Leptoquarks}

In LQ models one adds in their minimal version only one new field to the SM. Therefore, they are, with respect to their particle content, minimal models with chiral enhancement. In constructing these models one demands that the couplings to quarks and leptons respect SM gauge invariance, resulting in $5$ vector LQs and $5$ scalar LQs~\cite{Buchmuller:1986zs}. Therefore, in Eqs.~\eqref{c_R_general_Phi} and \eqref{c_R_general_V} $M_\Psi$ corresponds to the quark mass, $M_\Phi$ and $M_V$ to the LQ mass, respectively, and a factor $N_c=3$ has to be added to take into account the fact that quarks and LQs are colored.
 
\begin{figure}
	\centering
	\includegraphics[width=0.59\textwidth]{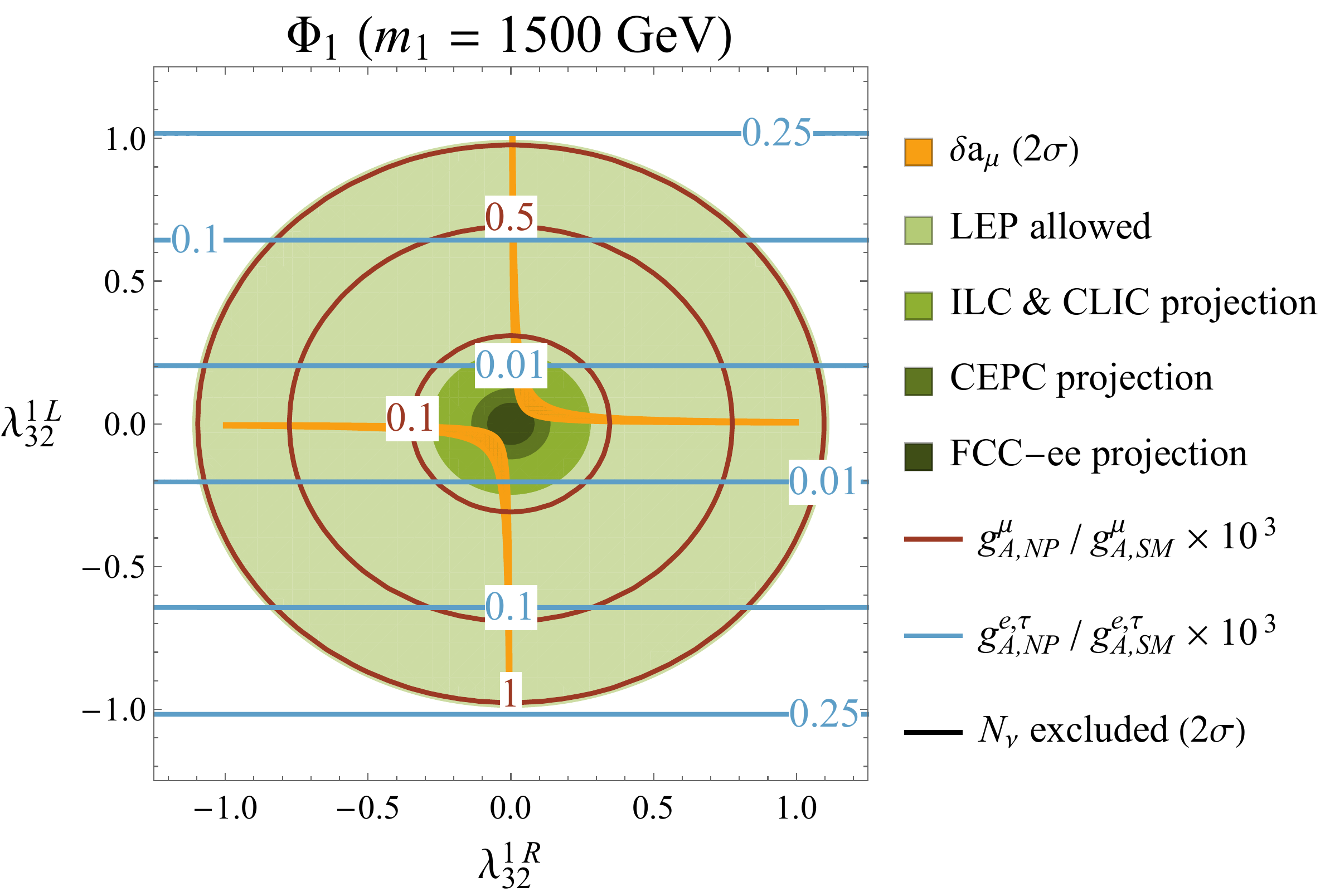}
	\includegraphics[width=0.395\textwidth]{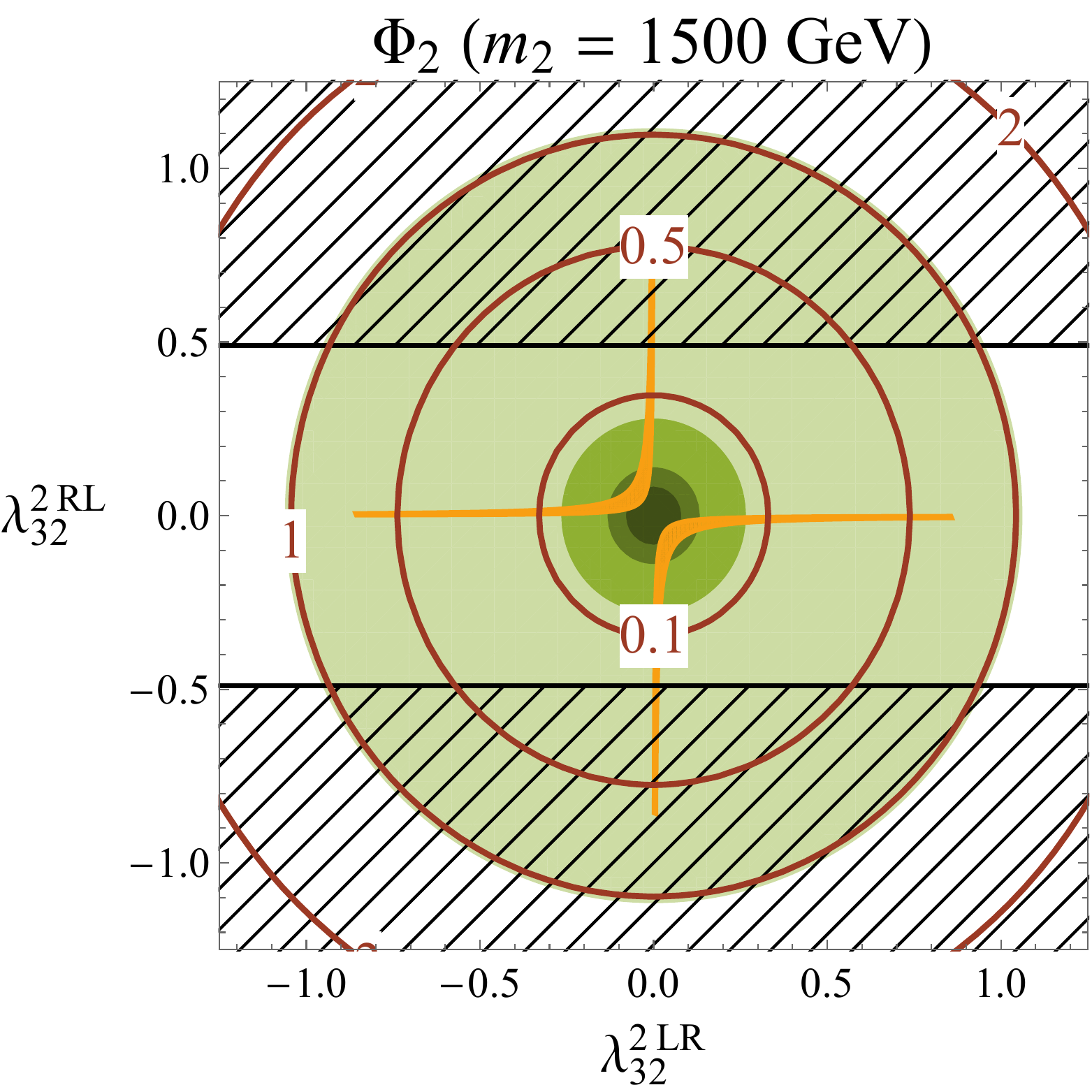}
	\caption{Allowed parameter space by LEP~\cite{ALEPH:2005ab} (light green) for the couplings to left- and right-handed muons. In addition, we give the expected sensitivities of future collider experiments. The finite renormalization of $g_{2}$, induced by the effect in the Fermi constant, yields a lepton-flavor-universal effect, which is depicted by the blue lines in the  left figure. Here the parameters $\lambda^{1(2)L}_{32}$ and $\lambda^{1(2)L}_{32}$ are the couplings of $S_1$ ($S_2$) to top quarks with left-handed and right-handed muons, respectively. Taken from Refs.~\cite{ColuccioLeskow:2016dox,Crivellin:2020mjs,Arnan:2019olv}.}
	\label{fig:Zmumu_AMM}
\end{figure}

There are two representations of scalar LQs that can easily accommodate $a_\mu$ via a chiral enhancement by the top mass of $m_t/m_\mu\approx 1600$~\cite{Djouadi:1989md,Davidson:1993qk,Couture:1995he,Chakraverty:2001yg,Bauer:2015knc,Das:2016vkr,Biggio:2016wyy,ColuccioLeskow:2016dox}, so that even for TeV-scale masses one can easily explain the tension in $a_\mu$. As in the generic case, sizable effects in $h\to\mu^+ \mu^-$ and $Z\to\mu^+\mu^-$ are observed~\cite{Arnan:2019olv}. However, the predictions are even more direct as $\kappa$ is given by the top Yukawa coupling. The predictions for $h\to\mu^+ \mu^-$ are given in Fig.~\ref{S1S2}, where the only free parameters are the trilinear Higgs--LQ couplings. Neglecting the effects of these couplings in $Z\to\mu^+\mu^-$, where they only lead to dim-8 effects, the predictions are shown in Fig.~\ref{fig:Zmumu_AMM}.

\section{Conclusions}

In these proceedings we reviewed the consequences of explaining $g-2$ of the muon in terms of heavy new physics. In practice, chiral enhancement is a necessary condition for such an explanation, otherwise, direct LHC mass limits exclude most of the available parameter space. In the general case, chiral enhancement requires at least three different fields in the loop, and we worked out the matching onto SMEFT as well as the resulting effects in $h\to\mu^+\mu^-$ and $Z\to \mu^+\mu^-$ for a wide range of such simplified models. Minimal scenarios can be constructed by using SM particles in the loop, e.g., if one identifies two of these fields with left- and right-handed top quarks, the third field becomes a scalar LQ, defining a minimal model with chiral enhancement. For such a concrete setup, the correlations with $h\to\mu^+ \mu^-$ and $Z\to\mu^+\mu^-$ become even more predictive. In this way, precision measurements of these processes at future colliders provide valuable complementary information on explanations of $a_\mu$ in terms of new degrees of freedom above the EW scale.

\acknowledgments

AC thanks the organizers, especially Gino Isidori, for the invitation to ``Les Rencontres de Physique de la Valle d'Aoste'' and the opportunity to present these results. This work is supported by the Swiss National Science Foundation, under Project Nos.\ PP00P21\_76884 and PCEFP2\_181117.

\end{document}